\documentstyle[emlines,bezier,floats,aps,amssymb,eqsecnum,preprint]{revtex}
\tighten

\newcommand{\ral}{\rangle}
\newcommand{\lal}{\langle}
\newcommand{\Img}{\mathop{\rm Im}}
\newcommand{\Real}{\mathop{\rm Re}}
\newcommand{\D}{\displaystyle}

\newcommand{\cD}{{\cal D}}
\newcommand{\cG}{{\cal G}}
\newcommand{\cH}{{\cal H}}

\newcommand{\re}{{\rm e}}
\newcommand{\ri}{{\rm i}}
\newcommand{\sP}{{\sf P}}
\newcommand{\sw}{{\sl w}}

\newcommand{\vphi}{{\varphi}}

\newcommand{\tM}{{\widetilde{M}}}

\newcommand{\reduction}[2]{\left.\phantom{\bigl|} #1 \right|_{#2}}
\newcommand{\C}{{\Bbb C}}
\newcommand{\R}{{\Bbb R}}
\newcommand{\T}{{\Bbb T}}
\newcommand{\Z}{{\Bbb Z}}
\begin{document}
\title{Perturbation of a lattice spectral band by a nearby 
resonance\thanks{LANL e-print cond-mat/9909414}}
\author{ A. K. Motovilov\thanks{On leave of absence from
         the Bogoliubov Laboratory of Theoretical Physics,
    Joint Institute  for Nuclear Research,
    141980 Dubna, Russia},
        W. Sandhas}
\address{Physikalisches Institut der Universit\"at Bonn,
Endenicher Allee 11-13, D-53115 Bonn, Germany}
\author{V. B. Belyaev}
\address{Joint Institute for Nuclear Research,
    141980 Dubna,  Russia}
\date{May 24, 2000}
\maketitle
\bigskip

\begin{abstract}
A soluble model of weakly coupled ``molecular'' and ``nuclear''
Hamiltonians is studied in order to exhibit explicitly the
mechanism leading to the enhancement of fusion probability in
case of a narrow near-threshold nuclear resonance. We, further,
consider molecular cells of this type being arranged in lattice
structures.  It is shown that if the real part of the narrow
nuclear resonance lies within the molecular band generated by
the intercellular interaction, an enhancement, proportional to
the inverse width of the nuclear resonance, is to be expected.
\bigskip


\end{abstract}

\section{Introduction}
\label{Intro}

Molecules are usually treated as purely Coulombic systems, while
the strong interaction between their nuclear constituents is
assumed to play a negligible role. However, at least in
principle, any Coulombic molecular level lying above the lower
threshold of the nuclear subsystem, is embedded in the
continuous spectrum of the nuclear sub-Hamiltonian. The coupling
between the molecular and nuclear channels, hence, turns this
level into a resonance (see, e.~g.,
Refs.~\cite{Albeverio72,Howland,Rauch,Hunziker,ReedSimonIV} and
references cited therein). Of course, due to the wide Coulombic
barrier between the nuclei and the short-range character of the
nuclear interaction, this coupling, and thus the width of the
resonance, which determines the fusion probability of the
nuclear constituents of the molecule, is in general extremely
small.

However, as pointed out in \cite{BelMotTMF,BelMotEprint}, the
situation is rather different if the nuclear subsystem of a
molecule has a sufficiently narrow near-threshold.  Examples of
such nuclear systems  may be read off from the data presented in
\cite{ENSDF}.  Among them are even customary systems like
$p\,p\,{}^{16}$O and
$p\,{}^{17}$O~\cite{83Ajzenberg-Selove18-20,95Levels18-19},
i.\,e., the nuclear constituents of the water molecule $H_2{O}$
or the hydroxyl ion $OH^-$ with $O$ being the isotope
${}^{17}$O.  For $Li D$ and $H_2{O}$ the influence of
near-threshold nuclear resonances on the molecular properties
has been studied in \cite{BMS1,BMS2,BMS3} by estimating the
overlap integrals between the corresponding molecular and
nuclear wave functions.  The best known example of such
phenomena is the muon catalyzed fusion of deuteron and triton in
the $dt\mu$ molecule, where the near-threshold nuclear resonance
${}^5$He$(3/2^+)$ plays a decisive role~\cite{BreunlichOthers}.

Being motivated by the above special cases, we deal in this paper
with a rather general model Hamiltonian related to the ones
considered by Friedrichs in~\cite{Friedrichs}.  This Hamiltonian
consists of a ``nuclear" part, a ``molecular'' part with
eigenvalues embedded in the continuous spectrum of the nuclear
part, and a weak coupling term which turns these unperturbed
eigenvalues into  ``molecular'' resonances.
Since the model is explicitly solvable, the mechanism
of formation of the resonances becomes clearly visible.

The following property pointed out
in~\cite{BelMotTMF,BelMotEprint} appears, in particular, as a
general feature:  if the ``nuclear'' channel itself has a narrow
resonance with a position close to the ``molecular'' energy,
then {\rm the width} (the imaginary part) {\rm of the resulting
``molecular'' resonance is found to be inversely proportional to
the ``nuclear'' width}.  In other words, a large increase of the
decay rate of the ``molecular'' state, i.\,e. of the fusion
probability, is observed in this case.  Such a coincidence of
nuclear and molecular energies is, of course, a rather rare
phenomenon in nature.

A further goal of the present work is to show that the decay rate
may be considerably enhanced when arranging molecular clusters
of this type within a crystalline structure. The reason is that
in such a configuration the original discrete molecular energy
turns into a band, i.\,e., into a whole interval of the
continuous spectrum%
\footnote{Note that in the models under consideration, the spectral band
generated by the ``molecular'' level  is shifted finally, after
switching on the coupling between the ``nuclear'' and
``molecular'' channels, into the unphysical sheet.}.
That is, even if the position of the ``nuclear'' resonance
differs from the original ``molecular'' level, it can get into this
band. This allows for a fine tuning by
exciting the crystalline structure to energies as close as
possible to the energy of the ``nuclear''
resonance.  We show that the lattice states, which correspond to
such an initial choice of their quasimomentum distribution,
decay exponentially with a rate which is again inversely
proportional to the width of the ``nuclear'' resonance.

The paper is organized as follows.

In Sec.~\ref{twochannel} we introduce the explicitly soluble
model designed to demonstrate the interplay of the molecular and
nuclear resonance widths.  It is also shown that in a wide time
interval the decay of the ``molecular'' state is indeed of the
standard exponential character~\cite{Baz'}.  This transition
will take place primarily into the open nuclear channels and its
rate is determined by the inverse width of the nuclear
resonance. In Sec.~\ref{Lattice1Dim} we consider the case where
molecular Hamiltonians of the type considered in
Sec.~\ref{twochannel} are arranged in form of an infinite
one-dimensional lattice.  Sec.~\ref{MultiDimensional} is devoted
to the generalization to multi-dimensional lattices. In both
these sections, the time evolution of originally pure molecular
states, extended to a spectral band within a lattice, is
studied. It is shown that if the real part of the ``nuclear''
resonance lies within such a spectral band, then there exist
molecular states which decay exponentially, with a rate
inversely proportional to the ``nuclear'' width.

\section{Two-channel molecular resonance model}
\label{twochannel}

\subsection{Description of the model Hamiltonian}
\label{TwoChannelDescr}

Let us consider a two-channel Hilbert space
$\cH=\cH_1\oplus\cH_2$ consisting of a ``nuclear''
Hilbert space $\cH_1$ (channel 1) and a one-dimensional
``molecular'' space $\cH_2=\C$ (channel 2).  The elements of
$\cH$ are represented as vectors
$
u=\left(\begin{array}{c}
       u_1 \\
       u_2
\end{array}\right)\,
$
where $u_1\in\cH_1$ and $u_2\in\cH_2$, with $u_2$ being a
complex number.  The inner product $\lal u,v\ral_{\cH}=\lal
u_1,v_1\ral +u_2\overline{v}_2$ in $\cH$ is naturally
defined via the inner products $\lal u_1,v_1\ral$ in
$\cH_1$ and $u_2\overline{v}_2$ in $\cH_2$.

As a Hamiltonian in $\cH$ we consider the $2\times2$
operator matrix
\begin{equation}
\label{H2}
A=\left(\begin{array}{cr}
h_1            & \quad b \\
\lal\,\cdot\,, b\ral   &        \lambda_2
\end{array}
\right)
\end{equation}
where $h_1$ is the (selfadjoint) ``nuclear Hamiltonian'' in
$\cH_1$, and $\lambda_2\in\R$  a trial ``molecular'' energy. A
vector $b\in\cH_1$ provides the coupling between the channels.
It should be mentioned that the Hamiltonian~(\ref{H2}) resembles
one of the well known Friedrichs models~\cite{Friedrichs}.

If there is no coupling between the channels, i.\,e. for
$b=0$, the spectrum of $A$ consists of the spectrum of
$h_1$ and the additional discrete eigenvalue $\lambda_2$.  We
assume that the continuous spectrum $\sigma_c(h_1)$ of the
Hamiltonian $h_1$ is not empty and that the eigenvalue
$\lambda_2$  is embedded in $\sigma_c(h_1)$.  It is also assumed
that $\lambda_2$ is not a threshold point of $\sigma_c(h_1)$,
and that this spectrum is absolutely continuous in a
sufficiently wide neighborhood of $\lambda_2$.

A nontrivial coupling between the channels will, in general,
shift the eigenvalue $\lambda_2$ into an unphysical sheet of the
energy plane. The resulting perturbed energy appears as a
resonance, i.\,e.,  as a pole of the analytic continuation of the
resolvent $r(z)=(A-z)^{-1}$  taken between suitable states (see,
e.\,g.,~\cite{ReedSimonIV}).  The explicit representation for
the resolvent $r(z)$ is easily seen to be
\begin{equation}
\label{H2resolv}
r(z)=\left(\begin{array}{ccc}
r_1(z)+\D\frac{r_1(z)b\lal\,\cdot\,,b\ral r_1(z)}{M_2(z)} &\, &
-\D\frac{r_1(z)b}{M_2(z)}\\
-\D\frac{\lal\,\cdot\,,b\ral r_1(z)}{M_2(z)} &\,& \D\frac{1}{M_2(z)}
\end{array}\right).
\end{equation}
Here $r_1(z)$ stands for the resolvent $r_1(z)=(h_1-z)^{-1}$ of
the ``nuclear'' Hamiltonian $h_1$, while the {transfer function}
$M_2(z)$ reads
$
M_2(z)=\lambda_2-z-\beta(z)
$
with  $\beta(z)=\lal r_1(z)b,b\ral$.  Evidently the poles of
$r(z)$ on the physical sheet are either due to zeros of the
function $M_2(z)$ or due to poles of the resolvent $r_1(z)$. The
latter correspond to the discrete spectrum of the operator $h_1$
which may determine part of the point spectrum of $A$. This is
true, in particular, for the multiple eigenvalues of $h_1$.  In
any case it is obvious that the perturbation of the eigenvalue
$\lambda_2$ only corresponds to solutions of the equation
$M_2(z)=0$, i.\,e., of
\begin{equation}
\label{H2res}
z=\lambda_2-\beta(z).
\end{equation}
This equation has no roots $z$ with $\Img z\neq0$ on the
physical sheet.  For, being eigenvalues of the selfadjoint
operator $A$, they have, of course, to be real. Thus, Eq.
(\ref{H2res}) may have solutions only on the real axis and in
the unphysical sheet(s) of the Riemann surface of the resolvent
$r_1(z)$.

We start with a brief discussion of the case where the
``nuclear'' channel Hamiltonian $h_1$ generates no resonances
close to $\lambda_2$ in a domain $\cD$ of the unphysical sheet
which ajoins the physical sheet from below the cut.  This
assumption implies that for a wide set of unit vectors
$\widehat{b}=b/\|b\|$ the quadratic form $\beta(z)=\|b\|^2\lal
r_1(z)\widehat{b},\widehat{b}\ral$ can be analytically continued
in $\cD$.  Moreover, under certain smallness conditions for
$\|b\|$, Eq.~(\ref{H2res}) is uniquely solvable~\cite{MennMot}
in $\cD$ providing in first order perturbation theory (see,
e.\,g.,~\cite{KMMM-YaF88,MotRemovalJMP})
\begin{equation}
\label{H2resonance}
z_2\mathop{=}\limits_{\|b\|\to 0}
\lambda_2-\lal r_1(\lambda_2+\ri0)b,b\ral
+o(\|b\|^2).
\end{equation}
The real and imaginary parts of the resonance
$z_2=E_R^{(2)}-\ri\D\frac{\Gamma_R^{(2)}}{2}$, thus, are given by
\begin{eqnarray}
\nonumber
E_R^{(2)} &=& \lambda_2-
\Real\lal r_1(\lambda_2+\ri0)b,b\ral+o(\|b\|^2), \\
\label{Gammaj}
\Gamma_R^{(2)} &=& 2\Img\lal r_1(\lambda_2+\ri0)b,b\ral
+o(\|b\|^2).
\end{eqnarray}

\subsection{Perturbation of the ``molecular'' resonance
by a nearby ``nuclear'' resonance}\label{Perturbation}

Our main interest concerns the opposite case of a
``nuclear'' resonance
$z_1=E_R^{(1)}-\ri\D\frac{\Gamma_R^{(1)}}{2}$,
$\Gamma_R^{(1)}>0$, with a real part $E_R^{(1)}$ close to
$\lambda_2$. For the sake of simplicity we assume the
corresponding pole of $r_1(z)$ to be of first order.  Let the
element $b\in\cH_1$ be such that the function $\beta(z)$ admits
an analytic continuation into a domain $\cD$ which contains both
points $\lambda_2$ and $z_1$. This domain, moreover, is assumed
to belong to the unphysical sheet which adjoins the physical
sheet along the upper rim of the cut.  In $\cD$ the function
$\beta(z)$, thus, can be written as
\begin{equation}
\label{BetaRepres}
\beta(z)=\D\frac{a}{z_1-z}+\beta^{\rm reg}(z)
\end{equation}
with $\beta^{\rm reg}(z)$ being a holomorphic function.  For a
fixed ``structure function'' \hbox{$\widehat{b}=b/\|b\|$} we
have $|a|=C_a\|b\|^2$ with a constant $C_a$ determined by the
residue of $r_1(z)$ at $z=z_1$.  Note that this residue is
usually expressed in terms of resonance (Gamow)
functions (see for example~\cite{MotMathNach}).  In fact, we
assume that the resonance corresponds to an ``almost
eigenstate'' of $h_1$. That is, in principle  a limiting
procedure $\Gamma_R^{(1)}\to 0$ is possible so that
the resonance turns into a usual eigenvalue with an eigenvector
$\psi_1\in\cH_1.$  More precisely, we assume
\begin{equation}
\label{alimit}
C_a= C_a^{(0)}+o(1) \quad \mbox{as}\quad {\Gamma_R^{(1)}\to 0}
\end{equation}
with $C_a^{(0)}\equiv\lal\widehat{b},
\psi_1\ral\lal\psi_1,\widehat{b}\ral\neq 0$.
This can be achieved, e.\,g.,
if the Hamiltonian $h_1$ itself has a
matrix representation of the form~(\ref{H2}) and the resonance
$z_1$ is generated by a separated one-dimensional channel.
In such a case, according to~(\ref{H2resolv}) and
(\ref{H2resonance}), we would have $C_a^{(0)}=1$
(for details see Ref.~\cite{BelMotEprint}, Sec.\,II).

Let
\begin{equation}
\label{ReaIma}
   \Real a>0 \quad\mbox{and}\quad\Img a\ll\Real a
\end{equation}
and, for $z\in\cD$,
$$
|\Img\beta^{\rm reg}(z)|\geq c_\cD\|b\|^2
\quad\mbox{and}\quad
|\beta^{\rm reg}(z)|\leq C_\cD \|b\|^2\,.
$$
with constants $c_\cD>0$ and $C_\cD>0$.  Furthermore,
the coupling between the channels in the
Hamiltonian~(\ref{H2}) is assumed to be so weak that
\begin{equation}
\label{H2Conditions}
\left|{\beta^{\rm reg}(z)}\right|\leq
C_\cD\|b\|^2\ll\Gamma_R^{(1)}\quad\mbox{while}\quad
|a|= C_a\|b\|^2\ll\left(\Gamma_R^{(1)}\right)^2.
\end{equation}
It can be expected that these conditions are fulfilled in
specific molecular systems even under the supposition that the
``nuclear'' width $\Gamma_R^{(1)}$ itself is very small.

After inserting~(\ref{BetaRepres}) for $\beta(z)$,
Eq.~(\ref{H2res}) turns into the ``quadratic'' equation
$$
(\lambda_2-z)(z_1-z)-a+(z_1-z)\beta^{\rm reg}(z)=0
$$
which can be ``solved'', i.\,e., can be rewritten in form of two
equations
\begin{equation}
\label{H2roots}
z=\D\frac{\lambda_2+z_1-\beta^{\rm reg}(z)}{2}\pm
\sqrt{\left(\D\frac{\lambda_2-z_1-\beta^{\rm reg}(z)}{2}\right)^2+a}.
\end{equation}
Under conditions~(\ref{H2Conditions}) the existence of
solutions of~(\ref{H2roots}), and thus of  Eq.~(\ref{H2res}),
is guaranteed, analogously to the proof
in~\cite{MennMot}, by Banach's Fixed Point
Theorem.  Each of the equations~(\ref{H2roots}) has only
one solution in the domain $\cD$.  In case of the sign ``$-$''
we denote the root of~(\ref{H2roots}) by $z_{\rm nucl}$,  in
case of the sign ``$+$'' by $z_{\rm mol}$.

The inequalities~(\ref{H2Conditions}) imply
\begin{equation}
\label{NeqEst2}
\D\frac{|a|}{  \left|\lambda_2-z_1-
\reduction{\beta^{\rm reg}(z)}{z\in\cD}\right|^2}  \approx
 \D\frac{|a|}{\left|\lambda_2-E_R^{(1)}\right|^2+
\left(\D\frac{ \Gamma_R^{(1)} }{ 2 }\right)^2}\leq
\D\frac{4C_a\|b\|^2}{ \left(\Gamma_R^{(1)}\right)^2 }\ll 1.
\end{equation}
For $z\in\cD$ the value of
\begin{equation}
\label{epsilon}
\varepsilon(z)=\D\frac{4a}{[\lambda_2-z_1-\beta^{\rm reg}(z)]^2}
\end{equation}
is very small, $|\varepsilon(z)|\ll 1$.
Thus, to separate the main terms of the solutions of
Eqs.~(\ref{H2roots}), one can apply the asymptotic relation
$\sqrt{1+\varepsilon}=1+\varepsilon/2+O(\varepsilon^2)$.  As a
result we find
\begin{equation}
\label{H2rootsAs}
z=\D\frac{\lambda_2+z_1-\beta^{\rm reg}(z)}{2}\pm
\D\frac{\lambda_2-z_1-\beta^{\rm reg}(z)}{2}
\left(1+\D\frac{2a}{(\lambda_2-z_1-\beta^{\rm reg}(z))^2}+
O(\varepsilon^2)\right).
\end{equation}
In other words, the roots
$z_{\rm nucl}$ and $z_{\rm mol}$ of~(\ref{H2roots})
are essentially given by
\begin{eqnarray}
\label{zNucl}
 z_{\rm nucl}  & \cong & z_1 -
\D\frac{a}{\lambda_2-z_1-\beta^{\rm reg}(z_1)}\cong
z_1-\D\frac{a}{ \lambda_2-z_1 },\\
\label{zMol}
z_{\rm mol} & \cong & \lambda_2-\beta^{\rm reg}(\lambda_2+\ri0)
+\D\frac{a}{\lambda_2-z_1-\beta^{\rm reg}(\lambda_2+\ri0)}\cong
\lambda_2+\D\frac{a}{\lambda_2-z_1}.
\end{eqnarray}
>From the second condition~(\ref{H2Conditions}) follows
$\left|\D\frac{a}{\lambda_2-z_1}\right|\ll \Gamma_R^{(1)}$.
Consequently, this term provides in $z_{\rm nucl}$ a very small
perturbation of the initial ``nuclear'' resonance $z_1$.  As
compared to $\Gamma_R^{(1)}$ it represents also in $z_{\rm mol}$
a very weak perturbation of the ``molecular'' energy
$\lambda_2$.  However, as compared to the
result~(\ref{H2resonance}), valid in case of a missing nearby
``nuclear'' resonance, it can be rather large.  In particular,
if the ``molecular'' energy $\lambda_2$ coincides with the real
part $E_R^{(1)}$ of the ``nuclear'' resonance $z_1$, then
$z_{\rm mol}=E_R^{(m)}-i\D\frac{\Gamma_R^{(m)}}{2}$ with
\begin{equation}
\label{GmFin}
E_R^{(m)}\cong\lambda_2-2\D\frac{\Img a}{\Gamma_R^{(1)}}
\quad\mbox{and}\quad
\Gamma_R^{(m)}\cong 4\D\frac{\Real a}{\Gamma_R^{(1)}}.
\end{equation}
{\it The width of the ``molecular'' resonance $z_{\rm mol}$ in
the presence of a nearby ``nuclear'' resonance $z_1$, thus,
turns out to be inversely proportional to the ``nuclear'' width
$\Gamma_R^{(1)}$}.

Let us contrast the results~(\ref{H2resonance})
and~(\ref{GmFin}) in some more detail.  Since such a comparison
is necessarily of a qualitative character, we simulate the
situation of a missing nearby nuclear resonance simply by
dropping the pole term in the representation~(\ref{BetaRepres})
of $\beta(z)$.  After this removal we get
$\beta(z)\equiv\beta^{\rm reg}(z)$ and for $\Img z\leq 0$ the
eigenvalue $\lambda_2$ generates the
resonance~(\ref{H2resonance}) having the width
$
\Gamma_R^{(2)}\approx
2\Img\beta^{\rm reg}(\lambda_2+\ri0).
$
The latter satisfies the inequalities
\mbox{$c_\cD\|b\|^2\leq\Gamma_R^{(2)}/2\leq C_\cD\|b\|^2.$}
Substituting $|\Real a|\sim C_a\|b\|^2\sim
\D\frac{C_a}{c_\cD}\Gamma_R^{(2)}$ in~(\ref{GmFin}) we find the
following approximate estimate of $\Gamma_R^{(m)}$ relative to
$\Gamma_R^{(2)}$:
\begin{equation}
\label{gmg2}
\Gamma_R^{(m)}\sim\Gamma_R^{(2)}\cdot
\D\frac{C_a/c_\cD}{\Gamma_R^{(1)}}.
\end{equation}

The second inequality~(\ref{H2Conditions}), chosen as a
condition for $\|b\|$ reflects the fact that the ``usual''
molecular width $\Gamma_R^{(2)}$ is much smaller than the width
of a usual ``nuclear'' resonance $\Gamma_R^{(1)}$,
\begin{equation}
\label{G2G1}
C_a\Gamma_R^{(2)}\ll  c_\cD\left(\Gamma_R^{(1)}\right)^2.
\end{equation}
This can practically always be assumed for concrete molecules.

Under condition~(\ref{alimit}) the value of $C_a=|a|/\|b\|^2$
differs from zero,  $C_a\geq C>0$, as $\Gamma_R^{(1)}\to 0$.
Therefore the estimates~(\ref{GmFin}) and~(\ref{gmg2}) imply
that in the presence of a narrow
($\Gamma_R^{(1)}\ll\D{C_a}/{c_\cD}$) ``nuclear'' resonance close
to $\lambda_2$ the ``molecular'' width $\Gamma_R^{(m)}$ is much
larger than the ``molecular'' width $\Gamma_R^{(2)}$ observed in
absence of such a resonance.  In fact, according to (\ref{gmg2})
this ratio is determined by the large quotient
$\D\frac{C_a/c_\cD}{\Gamma_R^{(1)}}$.

Finally we note that if the conditions~(\ref{H2Conditions}), and
hence the condition~(\ref{G2G1}) are not fulfilled, i.\,e., if the
coupling between the channels in the Hamiltonian~(\ref{H2}) is
not small compared with the ``nuclear'' width,
then it follows from~(\ref{H2roots}) that the
molecular width $\Gamma_R^{(m)}$ achieves itself an order of
$\Gamma_R^{(1)}$.  We do not discuss this case since such a
situation can hardly be assumed to exist.

\subsection{Exponential decay of the ``molecular'' state}
\label{DecayTwochannel}

Let us suppose that an initial
state of the system described by the Hamiltonian
$A$ corresponds exactly to the pure
``molecular'' wave function
$\varphi=\left(\begin{array}{c} 0\\1 \end{array} \right)$.
Then, the time evolution  of the system is
described by the solution $\psi(t)$ of the Cauchy problem
\begin{equation}
\label{EvolutionTwoChannel}
    i\D\frac{d\psi}{dt}=A\psi, \quad \reduction{\psi}{t=0}=\varphi.
\end{equation}
The probability of finding the system at the time
$t$ still in the molecular state $\vphi$ is given by
$$
P_{\rm mol}(t)=|\lal\psi(t),\vphi\ral|^2.
$$
The remainder $1-P_{\rm mol}(t)$, hence, determines the
probability for the state $\vphi$ to decay into open channels
of the continuous spectrum of the ``nuclear''
sub-Hamiltonian $h_1$.

To estimate the probability $P_{\rm mol}(t)$, we use the
standard integral representation of a function of an operator via its
resolvent.  In the case considered this means
\begin{equation}
\label{intexpA}
\exp{\{-\ri A t\}}=-\D\frac{1}{2\pi \ri}\D\oint\limits_\gamma
dz \,{\rm e}^{-{\rm i}zt}(A-z)^{-1}.
\end{equation}
The integration in~(\ref{intexpA}) is performed in the physical
sheet along a contour $\gamma$ going counterclockwise around the
spectrum of the matrix $A$. Recall that, due to the
selfadjointness of the operator $A$, this spectrum is real.
Taking into account the representations~(\ref{H2resolv})
and~(\ref{intexpA}) one finds
\begin{equation}
\label{ProbScalTwoChannel}
\lal\psi(t),\vphi\ral=-\D\frac{1}{2\pi\ri}\oint\limits_\gamma
dz \, \frac{\exp(-{\rm i}zt)}
{\,\lambda_2-z-\beta(z)\,}
\end{equation}

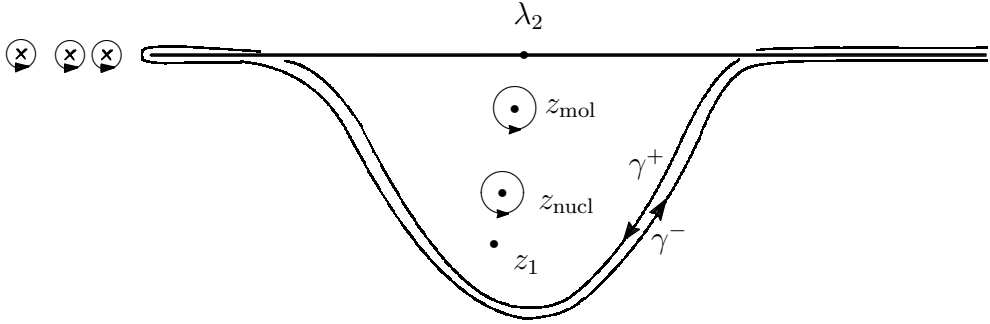
\begin{figure}
\centering
\unitlength=0.80mm
\special{em:linewidth .6pt}
\linethickness{.6pt}
\begin{picture}(159.33,76.00)
\emline{20.67}{69.67}{1}{159.33}{69.67}{2}
\emline{20.67}{69.33}{3}{159.33}{69.33}{4}
\bezier{164}(55.33,57.67)(69.00,30.33)(79.67,28.00)
\bezier{184}(53.67,55.67)(68.00,27.67)(82.67,25.67)
\put(79.00,46.66){\circle{7.33}}
\emline{79.87}{43.07}{5}{78.13}{43.58}{6}
\emline{78.16}{42.56}{7}{79.80}{43.07}{8}
\emline{78.13}{43.58}{9}{78.33}{43.07}{10}
\emline{78.33}{43.07}{11}{78.13}{42.49}{12}
\emline{78.20}{43.38}{13}{79.84}{43.11}{14}
\emline{79.84}{43.11}{15}{78.20}{42.76}{16}
\emline{78.27}{43.21}{17}{79.74}{43.11}{18}
\emline{78.30}{42.97}{19}{79.60}{43.11}{20}
\emline{78.10}{43.73}{21}{79.90}{43.09}{22}
\put(79.00,46.47){\circle*{1.48}}
\emline{20.44}{69.50}{23}{51.33}{69.50}{24}
\emline{51.36}{69.48}{25}{58.00}{69.48}{26}
\emline{58.07}{69.46}{27}{75.67}{69.46}{28}
\emline{75.70}{69.50}{29}{105.67}{69.50}{30}
\emline{105.65}{69.51}{31}{132.67}{69.51}{32}
\emline{132.50}{69.54}{33}{159.33}{69.54}{34}
\bezier{72}(55.39,57.68)(48.89,67.70)(42.70,68.52)
\bezier{104}(53.64,55.52)(48.17,67.08)(35.37,68.42)
\bezier{64}(38.57,70.07)(24.74,71.10)(23.09,70.90)
\bezier{52}(35.37,68.32)(27.32,68.32)(22.78,68.11)
\bezier{16}(23.09,70.89)(19.49,70.87)(19.25,70.44)
\bezier{16}(22.79,68.11)(19.31,68.37)(19.04,68.77)
\bezier{8}(19.06,68.75)(18.74,70.22)(19.23,70.44)
\emline{20.45}{69.50}{35}{20.67}{69.64}{36}
\emline{20.45}{69.50}{37}{20.65}{69.32}{38}
\bezier{52}(91.26,30.23)(87.57,26.54)(79.94,27.81)
\bezier{44}(92.40,29.09)(89.60,26.03)(82.74,25.65)
\bezier{124}(91.38,30.23)(101.18,38.75)(108.68,55.54)
\bezier{132}(111.35,55.29)(103.85,39.01)(92.53,28.96)
\bezier{60}(108.65,55.49)(112.92,64.81)(116.73,67.75)
\bezier{68}(120.62,67.66)(115.68,66.71)(111.40,55.30)
\bezier{60}(121.10,70.32)(124.99,70.79)(136.21,70.70)
\bezier{40}(134.78,70.70)(142.67,70.79)(144.76,70.70)
\bezier{52}(120.72,67.66)(124.61,68.61)(133.36,68.51)
\bezier{44}(133.36,68.51)(140.01,68.70)(144.76,68.61)
\bezier{8}(116.75,67.72)(117.38,68.25)(118.17,68.70)
\bezier{60}(144.79,70.69)(154.78,70.66)(159.33,70.73)
\bezier{60}(144.72,68.59)(156.03,68.56)(159.33,68.59)
\put(83.33,76.00){\makebox(0,0)[cc]{$\lambda_2$}}
\put(90.33,60.67){\makebox(0,0)[cc]{$z_{\rm mol}$}}
\emline{99.15}{38.81}{39}{99.99}{41.64}{40}
\emline{99.99}{41.64}{41}{100.56}{40.71}{42}
\emline{100.56}{40.71}{43}{101.69}{40.48}{44}
\emline{101.69}{40.48}{45}{99.18}{38.81}{46}
\emline{99.18}{38.81}{47}{100.15}{41.35}{48}
\emline{100.15}{41.35}{49}{100.37}{41.06}{50}
\emline{100.37}{41.06}{51}{99.12}{38.68}{52}
\emline{99.12}{38.68}{53}{100.92}{40.64}{54}
\emline{100.92}{40.64}{55}{101.27}{40.55}{56}
\emline{101.27}{40.55}{57}{99.15}{38.81}{58}
\put(89.66,44.67){\makebox(0,0)[cc]{$z_{\rm nucl}$}}
\put(83.00,34.67){\makebox(0,0)[cc]{$z_1$}}
\put(106.67,39.00){\makebox(0,0)[cc]{$\gamma^-$}}
\put(103.00,51.33){\makebox(0,0)[cc]{$\gamma^+$}}
\put(82.58,69.43){\circle*{1.48}}
\put(77.54,38.04){\circle*{1.48}}
\emline{78.09}{42.43}{59}{79.88}{43.09}{60}
\put(81.00,60.66){\circle{7.33}}
\emline{81.87}{57.07}{61}{80.13}{57.58}{62}
\emline{80.16}{56.56}{63}{81.80}{57.07}{64}
\emline{80.13}{57.58}{65}{80.33}{57.07}{66}
\emline{80.33}{57.07}{67}{80.13}{56.49}{68}
\emline{80.20}{57.38}{69}{81.84}{57.11}{70}
\emline{81.84}{57.11}{71}{80.20}{56.76}{72}
\emline{80.27}{57.21}{73}{81.74}{57.11}{74}
\emline{80.30}{56.97}{75}{81.60}{57.11}{76}
\emline{80.10}{57.73}{77}{81.90}{57.09}{78}
\put(81.00,60.47){\circle*{1.48}}
\emline{80.09}{56.43}{79}{81.88}{57.09}{80}
\emline{13.28}{69.52}{81}{12.68}{70.27}{82}
\emline{12.68}{70.27}{83}{13.72}{68.92}{84}
\emline{13.72}{68.92}{85}{13.20}{69.52}{86}
\emline{13.20}{69.52}{87}{14.10}{70.27}{88}
\emline{14.10}{70.27}{89}{12.38}{68.77}{90}
\emline{12.38}{68.77}{91}{13.87}{70.27}{92}
\emline{13.87}{70.27}{93}{13.13}{69.60}{94}
\emline{13.13}{69.60}{95}{12.53}{70.27}{96}
\emline{12.53}{70.27}{97}{13.87}{68.70}{98}
\put(13.20,69.52){\circle{4.78}}
\emline{14.41}{67.46}{99}{12.32}{67.96}{100}
\emline{12.32}{67.96}{101}{12.30}{66.61}{102}
\emline{12.30}{66.61}{103}{14.42}{67.42}{104}
\emline{14.42}{67.42}{105}{12.30}{67.73}{106}
\emline{12.30}{67.73}{107}{12.32}{67.37}{108}
\emline{12.32}{67.37}{109}{14.41}{67.42}{110}
\emline{14.41}{67.42}{111}{12.30}{67.10}{112}
\emline{12.30}{67.10}{113}{12.30}{66.82}{114}
\emline{12.30}{66.82}{115}{14.39}{67.42}{116}
\emline{14.39}{67.42}{117}{12.30}{67.55}{118}
\emline{7.19}{69.52}{119}{6.59}{70.27}{120}
\emline{6.59}{70.27}{121}{7.63}{68.92}{122}
\emline{7.63}{68.92}{123}{7.11}{69.52}{124}
\emline{7.11}{69.52}{125}{8.01}{70.27}{126}
\emline{8.01}{70.27}{127}{6.29}{68.77}{128}
\emline{6.29}{68.77}{129}{7.78}{70.27}{130}
\emline{7.78}{70.27}{131}{7.04}{69.60}{132}
\emline{7.04}{69.60}{133}{6.44}{70.27}{134}
\emline{6.44}{70.27}{135}{7.78}{68.70}{136}
\put(7.11,69.52){\circle{4.78}}
\emline{8.31}{67.46}{137}{6.23}{67.96}{138}
\emline{6.23}{67.96}{139}{6.21}{66.61}{140}
\emline{6.21}{66.61}{141}{8.33}{67.42}{142}
\emline{8.33}{67.42}{143}{6.21}{67.73}{144}
\emline{6.21}{67.73}{145}{6.23}{67.37}{146}
\emline{6.23}{67.37}{147}{8.31}{67.42}{148}
\emline{8.31}{67.42}{149}{6.21}{67.10}{150}
\emline{6.21}{67.10}{151}{6.21}{66.82}{152}
\emline{6.21}{66.82}{153}{8.30}{67.42}{154}
\emline{8.30}{67.42}{155}{6.21}{67.55}{156}
\emline{-0.93}{69.67}{157}{-1.53}{70.42}{158}
\emline{-1.53}{70.42}{159}{-0.49}{69.07}{160}
\emline{-0.49}{69.07}{161}{-1.01}{69.67}{162}
\emline{-1.01}{69.67}{163}{-0.11}{70.42}{164}
\emline{-0.11}{70.42}{165}{-1.83}{68.92}{166}
\emline{-1.83}{68.92}{167}{-0.34}{70.42}{168}
\emline{-0.34}{70.42}{169}{-1.08}{69.75}{170}
\emline{-1.08}{69.75}{171}{-1.68}{70.42}{172}
\emline{-1.68}{70.42}{173}{-0.34}{68.85}{174}
\put(-1.01,69.67){\circle{4.78}}
\emline{0.19}{67.60}{175}{-1.90}{68.11}{176}
\emline{-1.90}{68.11}{177}{-1.91}{66.75}{178}
\emline{-1.91}{66.75}{179}{0.21}{67.57}{180}
\emline{0.21}{67.57}{181}{-1.91}{67.88}{182}
\emline{-1.91}{67.88}{183}{-1.90}{67.52}{184}
\emline{-1.90}{67.52}{185}{0.19}{67.57}{186}
\emline{0.19}{67.57}{187}{-1.91}{67.24}{188}
\emline{-1.91}{67.24}{189}{-1.91}{66.97}{190}
\emline{-1.91}{66.97}{191}{0.17}{67.57}{192}
\emline{0.17}{67.57}{193}{-1.91}{67.70}{194}
\emline{99.08}{38.59}{195}{100.52}{40.75}{196}
\emline{100.52}{40.75}{197}{100.70}{40.63}{198}
\emline{100.70}{40.63}{199}{99.14}{38.65}{200}
\emline{104.59}{43.09}{201}{106.09}{45.43}{202}
\emline{106.09}{45.43}{203}{103.51}{43.33}{204}
\emline{103.51}{43.33}{205}{104.71}{43.15}{206}
\emline{104.71}{43.15}{207}{105.25}{42.01}{208}
\emline{105.25}{42.01}{209}{106.15}{45.49}{210}
\emline{106.15}{45.49}{211}{103.81}{43.27}{212}
\emline{103.81}{43.27}{213}{104.23}{43.27}{214}
\emline{104.23}{43.27}{215}{106.15}{45.43}{216}
\emline{106.15}{45.43}{217}{104.83}{42.97}{218}
\emline{104.83}{42.97}{219}{105.01}{42.67}{220}
\emline{105.01}{42.67}{221}{106.09}{45.43}{222}
\emline{106.09}{45.43}{223}{105.13}{42.37}{224}
\emline{104.05}{43.27}{225}{106.03}{45.31}{226}
\end{picture}
\vskip-1.5cm
\caption{A scheme showing the deformation of the integration
path $\gamma$.  The part $\gamma^+$ of the resulting contour
belongs to the unphysical sheet, part $\gamma^-$ to the physical
sheet. The crosses ``$\times$'' denote the discrete eigenvalues
of the Hamiltonian $A$ while the solid line corresponds to the
continuous spectrum.}
\label{fig-path:pic}
\end{figure}

This leads to the following important result.
Under the conditions of Subsection~\ref{Perturbation}
the behavior of the integral~(\ref{ProbScalTwoChannel}) for
$t>0$  is  described by the formula
\begin{eqnarray}
\label{ProbDetail}
\lal\psi(t),\vphi\ral &=& \exp\{-{\rm i}z_{\rm mol} t\}
\left[1-\D\frac{a}{\biggl(\lambda_2-z_1-
\beta^{\rm reg}(\lambda_2+\ri0)\biggr)^2} +
O\biggl(\varepsilon^4(\lambda_2+\ri0)\biggr)\right]  \\
\nonumber
 &+& \exp\{-{\rm i}z_{\rm nucl} t\}
\left[\D\frac{a}{\biggl(\lambda_2-z_1-
\beta^{\rm reg}(z_1)\biggr)^2} +
O\biggl(\varepsilon^4(z_1)\biggr)\right]
+\widetilde{\varepsilon}(t)
\end{eqnarray}
where the value of $\varepsilon(z)$ is given by Eq.~(\ref{epsilon}).
The background term
$\widetilde{\varepsilon}(t)=O{(\|b\|^2)}$ is small,
$|\widetilde{\varepsilon}(t)|\ll 1$, for all $t>0$.
In particular, if $E_R^{(1)}=\lambda_2$ we have
\begin{equation}
\label{ProbGamma}
\lal\psi(t),\vphi\ral \approx \exp\{-{\rm i}z_{\rm mol} t\}
\left[1+\D\frac{4a}{\biggl(\Gamma_R^{(1)}\biggr)^2} +\ldots\right]
 + \exp\{-{\rm i}z_{\rm nucl} t\}
\left[-\D\frac{4a}{\biggl(\Gamma_R^{(1)}\biggr)^2} +\ldots\right]
\end{equation}

The proof of the asymptotic relation~(\ref{ProbDetail}) is
carried out by estimating the contribution  of the resonance
poles $z_{\rm mol}$ and $z_{\rm nucl}$ to the
integral~(\ref{ProbScalTwoChannel}).  This is done by deforming
parts of the contour $\gamma$ situated in a neighborhood of the
molecular energy $\lambda_2$ (see Figure~\ref{fig-path:pic}). A
part $\gamma^+$ of $\gamma$, situated initially on the upper rim
of the cut, is shifted into the neighboring unphysical sheet.
Having done such a deformation one finds explicitly the residues
of the integrand in~(\ref{ProbScalTwoChannel}) at $z=z_{\rm
mol}$ and $z=z_{\rm nucl}$.  An analogous deformation of a part
$\gamma^-$ of $\gamma$, situated initially on the lower rim, is
performed in a domain $\Img z<0$ of the physical sheet.  It is
assumed that, though the parts $\gamma^+$ and $\gamma^-$ belong
to different energy sheets, their positions on these sheets
coincide.  It is also assumed that for any $z\in\gamma^\pm$ the
estimate $|\beta^\pm(z)|\ll|\lambda_2-z|$ holds.  Thus, the
integration in~(\ref{ProbScalTwoChannel}) around the continuous
spectrum of $A$, except the residues at $z=z_{\rm mol}$ and
$z=z_{\rm nucl}$, gives
\begin{eqnarray}
\nonumber
\lefteqn{ -\D\frac{1}{2\pi\ri}\int\limits_{\gamma^+}
dz \,\exp(-{\rm i}zt)\,\left( \frac{1}{\,\lambda_2-z-\beta^+(z)\,}-
\frac{1}{\,\lambda_2-z-\beta^-(z)\,}\right)\,}
\\
\label{IntGammaPm}
&=&-\D\frac{1}{2\pi\ri}\int\limits_{\gamma^+}
dz \,\exp(-{\rm i}zt)\,\,\frac{\beta^+(z)-\beta^-(z)\,}
{\,[\lambda_2-z-\beta^+(z)]\,[\lambda_2-z-\beta^-(z)]\,}.
\end{eqnarray}
Here  we have used the specific notation $\beta^+(z)$ for the
values of the function $\beta(z)$ at points $z$ belonging to the
curve $\gamma^+$ (i.\,e., lying in the unphysical sheet), and
$\beta^-(z)$ for the values of $\beta(z)$ at the same points of
the curve $\gamma^-$ (i.\,e., lying in the physical sheet). Both
$\beta^-(z)$ and $\beta^+(z)$ are of the order of $O(\|b\|^2)$,
and $|\beta^\pm(z)|\ll|\lambda_2-z|$, while the exponential
$\exp(-{\rm i}zt)$ at $\Img z<0$ is decreasing for $t>0$. The
value of the function~(\ref{IntGammaPm}), thus, is always small,
having an order of $O(\|b\|^2)$, and is even decreasing (in
general nonexponentially) with increasing $t$. We include the
contribution of this function in the background term
$\widetilde{\varepsilon}(t)$.  The summand
$\widetilde{\varepsilon}(t)$ also includes a contribution
to~(\ref{ProbScalTwoChannel}) from the residues at the discrete
eigenvalues of $A$.  Apart from factors oscillating
when $t$ changes, the value of this contribution remains
practically the same for all $t\geq0$.

The formulae~(\ref{ProbDetail}) and~(\ref{ProbGamma}) show
explicitly that { in a large time interval \mbox{$0\leq t<T$},
\mbox{$T\sim \D\frac{2}{{\Gamma_R^{(m)}}}\biggl|\ln\mathop{\rm
max}|\widetilde{\varepsilon}(t)|\biggr|$,} the decay of a
``molecular'' state $\vphi$ in the presence of a narrow
``nuclear'' resonance is indeed of exponential character.  The
rate of this decay is determined mainly by the width
$\Gamma_R^{(m)}$ of the ``molecular'' resonance $z_{\rm mol}$,
i.\,e., by the ratio $|\Real
a|/\Gamma_R^{(1)}$},
\begin{equation}
\label{Exponent}
P_{\rm mol}(t)\cong\exp\{-{\Gamma_R^{(m)}} t\}
\cong\exp\left\{-\frac{4|\Real a|}{\Gamma_R^{(1)}}\,t\right\}.
\end{equation}

\section{``Molecular'' resonances in a one-dimensional lattice}
\label{Lattice1Dim}

Let us assume that the ``molecules'' described by the
Hamiltonian~(\ref{H2}) are arranged in
form of an infinite one-dimensional (linear) crystalline
structure.  To describe such a crystal we introduce the
lattice Hilbert space
$$
\cG=\mathop{\oplus}\limits_{i=-\infty}^{+\infty}\cH^{(i)}
$$
representing an orthogonal sum of the Hilbert spaces
associated with the individual cells
\begin{equation}
\label{cHi}
\cH^{(i)}=\cH^{(i)}_1\oplus\cH^{(i)}_2\,.
\end{equation}
Here the subspaces $\cH^{(i)}_1\equiv\cH_1$ and
$\cH^{(i)}_2\equiv\cH_2\equiv\C$ are exactly the same ones as in
Sec.~\ref{twochannel} and, thus, $\cH^{(i)}\equiv\cH$.
The elements of the total Hilbert space $\cG$ are represented by the
sequences
$u=(\ldots,u^{(-2)},u^{(-1)},u^{(0)},u^{(1)},u^{(2)},\ldots)$
with components
$
u^{(i)}=\left(\begin{array}{c}
       u^{(i)}_1 \\
       u^{(i)}_2
\end{array}\right)
$
where $u^{(i)}_1\in\cH_1$ and $u^{(i)}_2\in\cH_2=\C$.
The inner product in $\cH$ is defined by
$
\lal u,v\ral_{\cH}=\sum\limits_{i=-\infty}^{+\infty}
\lal u^{(i)},v^{(i)}\ral_{\cH^{(i)}}.
$
The subspaces
$\cG_1=\mathop{\oplus}\limits_{i=-\infty}^{+\infty}\cH^{(i)}_1$
and
$\cG_2=\mathop{\oplus}\limits_{i=-\infty}^{+\infty}\cH^{(i)}_2$,
with  $\cG=\cG_1\oplus\cG_2$, are called pure nuclear and pure
molecular channels, respectively.

In the present section we deal with the Hamiltonian $H$ acting
in $\cH$ according to
\begin{equation}
\label{1dimLatHam}
(Hu)^{(i)}=Wu^{(i-1)}+Au^{(i)}+Wu^{(i+1)}
\end{equation}
where only the interaction between neighboring cells
is taken into account and the interaction operator
$W$ is chosen in the simplest form
\begin{equation}
\label{W1dim}
W=\left(\begin{array}{ccc}
       0              & & 0 \\
       0              & & {\sl w}
\end{array}\right)\,
\end{equation}
with ${\sl w}$ being a positive number.  Such a
choice of the interaction corresponds to the natural assumption
that the cells interact between each other via the molecular
states, while the direct interaction between nuclear constituents
belonging to different cells is negligible.  We assume that the
closed interval \mbox{$[\lambda_2-2{\sl w},\lambda_2+2{\sl w}]$}
is totally embedded in the continuous spectrum $\sigma_c(h_0)$
of $h_0$ and, moreover, that no thresholds of
$\sigma_c(h_0)$ belong to this interval.
For the sake of simplicity we also
assume that the interval belongs to the domain $\cD$ introduced
in Sec.~\ref{twochannel} and that for any
\mbox{$\mu\in[\lambda_2-2{\sl w},\lambda_2+2{\sl w}]$}
\begin{equation}
\label{NeverZero}
\Img\lal r_0(\mu\pm\ri0)\widehat{b},\widehat{b}\ral\neq0.
\end{equation}

Obviously, the Hamiltonian~(\ref{1dimLatHam}) represents a
special case of the infinite Jacobi operator matrix
(regarding the properties of some infinite  scalar Jacobi
matrices see, e.\,g., \cite{Klopp,GesztesyTeschl} and Refs.
cited therein).  It is a selfadjoint operator on the domain
$\cD(H)=\mathop{\oplus}\limits_{i=-\infty}^{+\infty}\cD^{(i)}$
with $\cD^{(i)}=\cD(h_1)\oplus\C$.

The resolvent $R(z)=(H-z)^{-1}$ of
$H$ possesses a natural block structure,
$R(z)=\{R(j,k;z)\}$, $j,k=0,\pm1,\pm2,...,\pm\infty$.
The blocks $R(j,k;z)$, providing the mapping
$\cH^{(k)}\to\cH^{(j)}$,
satisfy the equations
\begin{equation}
\label{Rjk}
WR(j-1,k;z)+(A-z)R(j,k;z)+WR(j+1,k;z)=\delta_{jk}I
\end{equation}
where $\delta_{jk}$ stands for the Kronecker delta and $I$ for
the identity operator in the Hilbert space $\cH$ of cells.
Hereafter we assume $\Img z\neq0$ so that the value of $z$
automatically belongs to the resolvent set of the operator $H$.
The blocks $R(j,k;z)$ themselves possess a $2\times2$ matrix
structure, $R(j,k;z)=\{R_{mn}(j,k;z)\}$, $m,n=1,2,$
corresponding to the decomposition
\mbox{$\cH=\cH_1\oplus\cH_2$}.

The Fourier transform
\begin{equation}
\label{Fourier}
(Fu)(p)=\frac{1}{\sqrt{2\pi}}\sum_{j=-\infty}^{+\infty}
    u^{(j)}\,\exp(\ri pj)\,
\end{equation}
in $\cG$ reduces Eq.~(\ref{Rjk}) to
\begin{equation}
\label{Rpp}
(A-z)R(p,p';z)+2\cos p\,WR(p,p';z)=\delta(p-p')\,I\,
\end{equation}
where the quasi-momentum $p$ runs through the interval
$[-\pi,\pi]$ and the function $R(p,p';z)$ represents the kernel
of the resolvent $R(z)$ in this representation. From~(\ref{Rpp}) follows
immediately
\begin{equation}
\label{RFactor}
R(p,p';z)=G(p;z)\delta(p-p')
\end{equation}
where
\begin{equation}
\label{Gpz}
G(p;z)=\left(\begin{array}{ccc}
r_1(z)+\D\frac{r_1(z)b\lal\,\cdot\,,b\ral r_1(z)}{\tM_2(p;z)} &\, &
-\D\frac{r_1(z)b}{\tM_2(p;z)}\\
-\D\frac{\lal\,\cdot\,,b\ral r_1(z)}{\tM_2(p;z)}
&\,& \D\frac{1}{\tM_2(p;z)}
\end{array}\right).
\end{equation}
This corresponds to the representation~(\ref{H2resolv})
of the resolvent of the cell Hamiltonian $A$,
the only difference being that the transfer function $M_2(z)$
is replaced by the expression
\begin{equation}
\label{TrFunctModif}
\tM_2(p;z)=\lambda_2-z+2{\sl w}\cos p-\beta(z)\,.
\end{equation}
The factorization~(\ref{RFactor}) implies
$$
   R(j,k;z)=\frac{1}{2\pi}
   \int_{-\pi}^{\pi}dp\,\re^{-\ri p(j-k)}\,G(p;z)\,
$$
and with the representation~(\ref{Gpz}) we, thus, obtain
\begin{equation}
\label{RjkExplicit}
R(j,k;z)=\left(\begin{array}{ccc}
\delta_{jk} r_1(z)+
 r_1(z)b\, R_{22}(j,k;z)\,\lal\,\cdot\,,b\ral r_1(z)  &\,\,\, &
- r_1(z)b\, R_{22}(j,k;z)\\
- R_{22}(j,k;z)\,\lal\,\cdot\,,b\ral r_1(z)
&\,& R_{22}(j,k;z)
\end{array}\right)
\end{equation}
where
\begin{equation}
\label{R22jk}
   R_{22}(j,k;z)=
  \frac{1}{2\pi}\int_{-\pi}^{\pi}dp\,\D\frac{\re^{-\ri p(j-k)}}
   {\lambda_2-z+2{\sl w}\cos p-\beta(z)}\,.
\end{equation}
Introducing the new variable $\zeta=\re^{-\ri p}$,
this integral is reduced to
$$
   R_{22}(j,k;z)=
  \frac{1}{2\pi\ri}\oint\limits_\gamma d\zeta\,\D\frac{\zeta^{j-k}}
   {\,\,{\sl w}\zeta^2+M_2(z)\zeta+{\sl w}\,\,}\,.
$$
Here, $\gamma$ stands for the unit circle centered
at the origin, the integration over $\gamma$ being performed in
the counterclockwise sense. Further, applying the Residue
Theorem and taking into account the sign of the difference
$j-k$ one finds
\begin{equation}
\label{R22jkz}
   R_{22}(j,k;z)=\D\frac{\left\{\D\frac{1}{2{\sl w}}
   \left[\sqrt{[M_2(z)-2{\sl w}][M_2(z)+2{\sl w}]}-M_2(z)
   \right]\right\}^{|j-k|}}
   {\sqrt{[M_2(z)-2{\sl w}][M_2(z)+2{\sl w}]}} \,.
\end{equation}
It is assumed here that the branch $\sqrt{(\xi-2{\sl
w})(\xi+2{\sl w})}$ of the function $\biggl((\xi-2{\sl
w})(\xi+2{\sl w})\biggr)^{1/2}$ is defined in the plane of the
complex parameter $\xi$, cut along the interval $[-2{\sl
w},2{\sl w}]$, and that $\Img\xi>0$ implies
$\Img\sqrt{(\xi-2{\sl w})(\xi+2{\sl w})}>0$, while $\Img\xi<0$
implies $\Img\sqrt{(\xi-2{\sl w})(\xi+2{\sl w})}<0$.

\begin{figure}
\centering
\unitlength=0.80mm
\special{em:linewidth .6pt}
\linethickness{.6pt}
\begin{picture}(159.33,76.00)
\emline{20.67}{69.67}{1}{159.33}{69.67}{2}
\emline{20.67}{69.33}{3}{159.33}{69.33}{4}
\emline{20.44}{69.50}{5}{51.33}{69.50}{6}
\emline{51.36}{69.48}{7}{58.00}{69.48}{8}
\emline{58.07}{69.46}{9}{75.67}{69.46}{10}
\emline{75.70}{69.50}{11}{105.67}{69.50}{12}
\emline{105.65}{69.51}{13}{132.67}{69.51}{14}
\emline{132.50}{69.54}{15}{159.33}{69.54}{16}
\emline{20.45}{69.50}{17}{20.67}{69.64}{18}
\emline{20.45}{69.50}{19}{20.65}{69.32}{20}
\put(83.33,76.00){\makebox(0,0)[cc]{$\lambda_2$}}
\put(79.00,34.34){\makebox(0,0)[cc]{$z_1$}}
\put(82.58,69.43){\circle*{1.48}}
\put(73.54,37.71){\circle*{1.48}}
\emline{13.28}{69.52}{21}{12.68}{70.27}{22}
\emline{12.68}{70.27}{23}{13.72}{68.92}{24}
\emline{13.72}{68.92}{25}{13.20}{69.52}{26}
\emline{13.20}{69.52}{27}{14.10}{70.27}{28}
\emline{14.10}{70.27}{29}{12.38}{68.77}{30}
\emline{12.38}{68.77}{31}{13.87}{70.27}{32}
\emline{13.87}{70.27}{33}{13.13}{69.60}{34}
\emline{13.13}{69.60}{35}{12.53}{70.27}{36}
\emline{12.53}{70.27}{37}{13.87}{68.70}{38}
\emline{7.19}{69.52}{39}{6.59}{70.27}{40}
\emline{6.59}{70.27}{41}{7.63}{68.92}{42}
\emline{7.63}{68.92}{43}{7.11}{69.52}{44}
\emline{7.11}{69.52}{45}{8.01}{70.27}{46}
\emline{8.01}{70.27}{47}{6.29}{68.77}{48}
\emline{6.29}{68.77}{49}{7.78}{70.27}{50}
\emline{7.78}{70.27}{51}{7.04}{69.60}{52}
\emline{7.04}{69.60}{53}{6.44}{70.27}{54}
\emline{6.44}{70.27}{55}{7.78}{68.70}{56}
\emline{-0.93}{69.67}{57}{-1.53}{70.42}{58}
\emline{-1.53}{70.42}{59}{-0.49}{69.07}{60}
\emline{-0.49}{69.07}{61}{-1.01}{69.67}{62}
\emline{-1.01}{69.67}{63}{-0.11}{70.42}{64}
\emline{-0.11}{70.42}{65}{-1.83}{68.92}{66}
\emline{-1.83}{68.92}{67}{-0.34}{70.42}{68}
\emline{-0.34}{70.42}{69}{-1.08}{69.75}{70}
\emline{-1.08}{69.75}{71}{-1.68}{70.42}{72}
\emline{-1.68}{70.42}{73}{-0.34}{68.85}{74}
\emline{62.32}{70.88}{75}{62.32}{67.94}{76}
\emline{62.37}{70.88}{77}{62.37}{67.94}{78}
\emline{62.42}{70.88}{79}{62.42}{67.94}{80}
\emline{102.51}{70.88}{81}{102.51}{67.94}{82}
\emline{102.56}{70.88}{83}{102.56}{67.94}{84}
\emline{102.61}{70.88}{85}{102.61}{67.94}{86}
\bezier{180}(62.33,61.67)(73.67,53.67)(102.67,64.67)
\bezier{44}(71.33,41.67)(73.67,44.00)(81.00,40.33)
\put(94.91,55.10){\makebox(0,0)[cc]{$z_{\rm mol}(p)$}}
\put(86.79,45.46){\makebox(0,0)[cc]{$z_{\rm nucl}(p)$}}
\put(104.00,76.00){\makebox(0,0)[cc]{$\lambda_2+2{\sl w}$}}
\put(61.33,75.88){\makebox(0,0)[cc]{$\lambda_2-2{\sl w}$}}
\bezier{180}(62.33,61.56)(73.67,53.56)(102.67,64.56)
\bezier{180}(62.33,61.33)(73.67,53.33)(102.67,64.33)
\bezier{44}(71.33,41.78)(73.67,44.11)(81.00,40.44)
\bezier{44}(71.33,41.78)(73.67,44.11)(81.00,40.44)
\bezier{44}(71.33,41.90)(73.67,44.23)(81.00,40.56)
\end{picture}
\vskip-1.5cm
\caption{A scheme showing the position of the resonance
bands generated in the unphysical sheet by the ``molecular''
eigenvalue $\lambda_2$ and the ``nuclear'' resonance $z_1$.
These bands are generated respectively by the points $z_{\rm
mol}(p)$ and $z_{\rm nucl}(p)$ with the quasimomentum $p$
running through the interval $[-\pi,\pi]$.}
\label{bands:pic}
\end{figure}
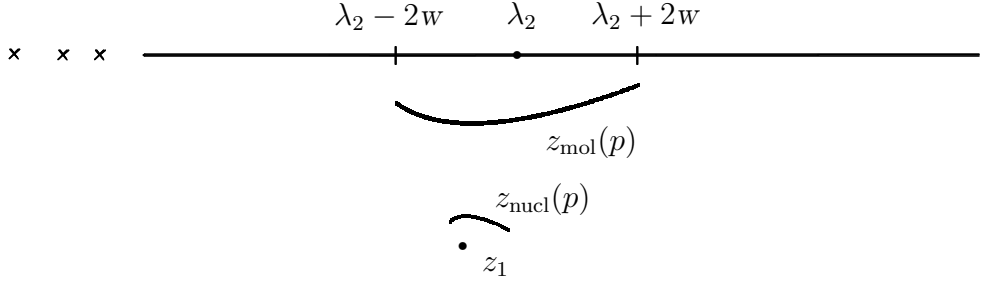

>From Eqs.~(\ref{RjkExplicit}) and~(\ref{R22jkz}) it follows
that all the nontrivial singularities of the resolvent $R(z)$,
differing from those of the cell ``nuclear'' channel
resolvent $r_1(z)$, are determined by the properties
of the function
$$
    D(z)=\sqrt{[M_2(z)-2{\sl w}][M_2(z)+2{\sl w}]}.
$$
First, we note that if $\|b\|=0$, and thus $M_2(z)=\lambda_2-z$,
then the ``molecular'' and ``nuclear'' channels in the
Hamiltonian $H$ decouple. In this case the eigenvalue
$\lambda_2$ generates for $H$ an additional branch of the
absolutely continuous spectrum which occupies the interval
\mbox{$[\lambda_2-2{\sl w},\lambda+2{\sl w}]$.} Second, even if
$\|b\|\neq0$ then the function $D(z)$ cannot have roots $z$
with \mbox{$\Img z\neq0$} in the physical sheet.  Otherwise such
roots would generate for $H$ a complex spectrum.  But this is
impossible because of the selfadjointness of $H$.  Also, under the
condition~(\ref{NeverZero}) this function cannot have real
roots within the interval $[\lambda_2-2{\sl w},\lambda_2+2{\sl
w}]$ since for $\lambda_2-2{\sl w}\leq\mu\leq\lambda_2+2{\sl w}$
the imaginary part
$$
\Img[M_2(\mu\pm\ri0)-2{\sl w}]=\Img[M_2(\mu\pm\ri0)+2{\sl w}]=
-\|b\|^2\Img\lal r_0(\mu\pm\ri0)\widehat{b},\widehat{b}\ral
$$
is nonzero. Thus, in a close neighborhood of the interval
$[\lambda_2-2{\sl w},\lambda_2+2{\sl w}]$ the equation $D(z)=0$
may only have roots in the unphysical sheet.  In fact, assuming
the conditions~(\ref{H2Conditions}) and repeating literally
the considerations which led to
(\ref{H2rootsAs}), one can rewrite this relation in
form of the four equations,
\begin{eqnarray}
\label{DLatRootMin}
z&=&\D\frac{\lambda_2-2{\sl w}+z_1-\beta^{\rm reg}(z)}{2}\\
\nonumber
&&\pm\D\frac{\lambda_2-2{\sl w}-z_1-\beta^{\rm reg}(z)}{2}
\left[1+\D\frac{2a}{(\lambda_2-2{\sl w}-z_1-\beta^{\rm reg}(z))^2}+
O(\varepsilon_-^2)\right]\,,\\
z&=&\D\frac{\lambda_2+2{\sl w}+z_1-\beta^{\rm reg}(z)}{2}\\
\nonumber
&&\pm\D\frac{\lambda_2+2{\sl w}-z_1-\beta^{\rm reg}(z)}{2}
\left[1+\D\frac{2a}{(\lambda_2+2{\sl w}-z_1-\beta^{\rm reg}(z))^2}+
O(\varepsilon_+^2)\right]\,,
\end{eqnarray}
where $\varepsilon_\pm=4a/(\lambda_2\pm 2{\sl w}-z_1-\beta^{\rm
reg}(z))$.  In that part of the domain $\cD$ which
belongs to the unphysical sheet, equation $D(z)=0$  has
four solutions being given essentially by
\begin{eqnarray}
\label{zNuclPM}
 z^{(\pm)}_{\rm nucl}  & \cong & z_1-
\D\frac{a}{\lambda_2\pm 2{\sl w}-z_1-\beta^{\rm reg}(z_1)}\cong
z_1-\D\frac{a}{ \lambda_2\pm 2{\sl w}-z_1 },\\
\label{zMolPM}
z^{(\pm)}_{\rm mol} & \cong & \lambda_2\pm 2{\sl w}-
\beta^{\rm reg}(\lambda_2\pm 2{\sl w}+\ri0)
+\D\frac{a}{\lambda_2\pm 2{\sl w}-z_1-
\beta^{\rm reg}(\lambda_2\pm 2{\sl w}+\ri0)} \\
\nonumber
   &\cong&\lambda_2\pm 2{\sl w}+\D\frac{a}{\lambda_2\pm 2{\sl w}-z_1}.
\end{eqnarray}
Obviously, each of the roots  $z^{(\pm)}_{\rm nucl}$ and
$z^{(\pm)}_{\rm mol}$ represents an additional square-root
branching point of the Riemann surface of the functions
$R_{22}(j,k;z)$.  Consequently, these roots are also the
branching points of the Riemann surface of the total resolvent
$R(z)$. Thus, one has to introduce the ``resonance'' cuts in the
unphysical sheet considered. The cuts can be made, say, between
$z^{(-)}_{\rm nucl}$ to $z^{(+)}_{\rm nucl}$ and between
$z^{(-)}_{\rm mol}$ and $z^{(+)}_{\rm mol}$. Evidently, these cuts
are to be interpreted as the resonance spectral bands generated
by the initial ``molecular'' level $\lambda_2$ and the ``nuclear''
resonance $z_1$  (see Figure~\ref{bands:pic}).

Consider now the time evolution of the system described by the
Hamiltonian $H$ starting from a pure molecular state
$\varphi=\varphi_1\oplus\varphi_2$, $\|\varphi_n\|\in\cG_n$,
$n=1,2$, with $\varphi_1=0$ and $\|\varphi\|=\|\varphi_2\|=1$.
The probability to find the system at a time
$t\geq0$ in the molecular channel is given by
\begin{equation}
\label{Pphi}
     P_{\rm mol}(\varphi,t)=\|\sP_2\re^{-\ri Ht}\varphi\|^2
\end{equation}
where $\sP_2$ is the orthogonal projection in $\cG$
on the pure molecular subspace $\cG_2$.

As in~(\ref{intexpA}) we represent the time evolution
operator $\exp(-\ri Ht)$ in terms of the
resolvent $R(z)=(H-z)^{-1}$,
\begin{equation}
\label{intexp}
\exp{\{-\ri Ht\}}=-\D\frac{1}{2\pi\ri}\D\oint\limits_\gamma
dz \,{\rm e}^{-\ri zt}(H-z)^{-1}
\end{equation}
where the integration is performed along a counterclockwise
contour $\gamma$ in the physical sheet
encircling the spectrum of the Hamiltonian $H$. Recall that this
spectrum is real since $H$ is a selfadjoint operator.

According to Eqs.~(\ref{RFactor}) and~(\ref{TrFunctModif}) the
operator $\reduction{\sP_2 R(z)}{\cG_2}$ acts in quasi-momentum
representation as the multiplication operator
\begin{equation}
\label{G22}
G_{22}(p,z)=[\tM_2(p;z)]^{-1}
\end{equation}
and, thus,
$$
   \biggl(\sP_2 R(z)\varphi\biggr)(p)=
    \D\frac{1}{\tM_2(p;z)}\,\varphi_2(p)\,, \quad p\in[-\pi,\pi]\,.
$$
Here $\varphi_2(p)$ stands for the values of the
Fourier transform~(\ref{Fourier}) of the vector
$\varphi_2=(\ldots,\varphi_2^{(-2)},\varphi_2^{(-1)},
\varphi_2^{(0)},\varphi_2^{(1)},\varphi_2^{(2)},\ldots)$,
which means
\begin{equation}
\label{P2phiFactor}
   \biggl(\sP_2\re^{-\ri Ht}\varphi\biggr)(p)=
       - \D\frac{1}{2\pi\ri}\,\, \varphi_2(p)\,J(p,t)
\end{equation}
with
\begin{equation}
\label{Jpt}
J(p,t)=\D\oint\limits_\gamma dz \,\D\frac{\exp(-\ri zt)}
       {\,\widetilde{\lambda}_2(p)-z-\beta(z)\,}\,.
\end{equation}
This expression has exactly the same form
as the integral~(\ref{ProbScalTwoChannel}).
The only difference consists in the replacement
of $\lambda_2$ by the sum
\begin{equation}
\label{lambdatilde}
\widetilde{\lambda}_2(p)=\lambda_2+2{\sl w}\cos p\,.
\end{equation}
Thus, to estimate the function $J(p,t)$ one
can immediately use the relation~(\ref{ProbDetail})
in order to find
\begin{eqnarray}
\nonumber
J(p,t) &=& \exp\{-{\rm i}z_{\rm mol}(p) t\}
\left[1-\D\frac{a}{\biggl(\widetilde{\lambda}_2(p)-z_1-
\beta^{\rm reg}(\widetilde{\lambda}_2(p)+\ri0)\biggr)^2} +
O\biggl(\varepsilon^4(p,\widetilde{\lambda}_2(p)+\ri0)\biggr)\right]  \\
\label{JptBehavior}
 &+& \exp\{-{\rm i}z_{\rm nucl}(p) t\}
\left[\D\frac{a}{\biggl(\widetilde{\lambda}_2(p)-z_1-
\beta^{\rm reg}(z_1)\biggr)^2} +
O\biggl(\varepsilon^4(p,z_1)\biggr)\right]
+\widetilde{\varepsilon}(p,t)\,,
\end{eqnarray}
where
\begin{equation}
\label{epsilonpz}
\varepsilon(p,z)=\D\frac{a}
{[\widetilde{\lambda}_2(p)-z_1-\beta^{\rm reg}(z)]^2}\,.
\end{equation}
The function $\widetilde{\varepsilon}(p,t)=O(\|b\|^2)$ is always
small, $|\widetilde{\varepsilon}(p,t)|\ll1$.  In accordance with
Eqs.~(\ref{zNucl}) and~(\ref{zMol}) we, hence, obtain for the positions
of the poles
\begin{eqnarray}
\label{zNuclp}
 z_{\rm nucl}(p)  & \cong &
z_1-\D\frac{a}{ \lambda_2+2{\sl w}\cos p-z_1 },\\
\label{zMolp}
z_{\rm mol}(p) & \cong &
\lambda_2+2{\sl w}\cos p+\D\frac{a}{\lambda_2+2{\sl w}\cos p-z_1}.
\end{eqnarray}
The resonance bands representing the ranges of the functions
$z_{\rm nucl}(p)$ and $z_{\rm mol}(p)$, with $p$ running through
the interval $[-\pi,\pi]$, are schematically depicted in
Figure~\ref{bands:pic}.

The asymptotics~(\ref{JptBehavior}) implies
\begin{equation}
\label{IntExpGamma}
P_{\rm mol}(\varphi,t)=\int_{-\pi}^{\pi}
dp\,|J(p,t)|^2\,|\varphi_2(p)|^2=
\int_{-\pi}^{\pi} dp\,
\exp\{-{\Gamma_R^{(m)}(p)}\,t\}\,|\varphi_2(p)|^2
+\widetilde{\varepsilon}(t)\,
\end{equation}
where
\begin{equation}
\label{GammaRm}
\Gamma_R^{(m)}(p)=
-2\Img z_{\rm mol}(p)\cong
-2\Img\D\frac{a}{\lambda_2+2{\sl w}\cos p-z_1}.
\end{equation}
The background term $\widetilde{\varepsilon}(t)$
in~(\ref{IntExpGamma}) is small for any $t\geq 0$,
$\widetilde{\varepsilon}(t)=O(\|b\|^2)$ and
$|\widetilde{\varepsilon}(t)|\ll1$.

Further, let us assume that the real part $E_R^{(1)}$ of the
``nuclear'' resonance $z_1$ belongs to the interval
$[\lambda_2-2{\sl w},\lambda_2+2{\sl w}]$, that is
$|E_R^{(1)}-\lambda_2|\leq2{\sl w}$. Then, one can always
prepare an initial ``molecular''
state $\varphi$ which decays via the ``nuclear'' channel with a
rate as close as possible to the desired maximal value.
Under the assumption~(\ref{ReaIma}), this maximum
is given by
$$
\mathop{\rm max}
\limits_{-\pi\leq p\leq\pi}
\Gamma_R^{(m)}(p)\cong 4\D\frac{\Real a}{\Gamma_R^{(1)}}
$$
(cf. formula~(\ref{GmFin})). The correspondingly prepared ``molecular''
state $\varphi$ has an almost monochromatic
component $\varphi_2(p)$
being localized in a close neighborhood of the
quasi-momenta
$$
  p=\pm\mathop{\rm arccos}\D\frac{E_R^{(1)}-\lambda_2}{2{\sl w}}\,.
$$
For example, if the function $\varphi_2(p)$ is nonzero only for
quasimomenta $p$ restricted by \mbox{$\left|\cos
p-\D\frac{E_R^{(1)}-\lambda_2}{2{\sl w}}\right|
\leq\delta\,\D\frac{\Gamma_R^{(1)}}{4{\sl w}}$} with some small
$\delta>0$, then the width $\Gamma_R^{(m)}$ given by the
relation~(\ref{GammaRm}) varies in an interval lying
approximately between $\D\frac{1}{1+\delta^2}\, \D\frac{4\Real
a}{\Gamma_R^{(1)}}$ and $\D\frac{4\Real a}{\Gamma_R^{(1)}}$.

\section{``Molecular'' resonances in a multi-dimensional lattice}
\label{MultiDimensional}

In this section we consider the case where
the ``molecules'' described by the Hamiltonians~(\ref{H2})
form an infinite $N$-dimensional crystalline structure.
To label the cells of the respective lattice we
use the multi-index $i\in\Z^N$, i.\,e.,
$i=(i_1,i_2,...,i_N)$ with $i_k=...,-2,-1,0,1,2,...$,
$k=1,2,...,N$. The Hilbert space of the system considered
is in this case
$
\cG=\mathop{\oplus}\limits_{i\in\Z^N}\cH^{(i)},
$
where the individual cell spaces
are given by (\ref{cHi}), with
$\cH^{(i)}_1\equiv\cH_1$ and
$\cH^{(i)}_2\equiv\cH_2\equiv\C$
being the spaces introduced in Sec.~\ref{twochannel}.
For the components $u^{(i)}\in\cH^{(i)}$ of the elements $u$
of the total Hilbert space $\cG$ we again use the column
representation
$
u^{(i)}=\left(\begin{array}{c}
       u^{(i)}_1 \\
       u^{(i)}_2
\end{array}\right)
$
with $u^{(i)}_1\in\cH_1$ and $u^{(i)}_2\in\cH_2=\C$.
The inner product in $\cH$ is defined as
$
\lal u,v\ral_{\cH}=\sum\limits_{i\in\Z^N}
\lal u^{(i)},v^{(i)}\ral_{\cH^{(i)}}.
$
The subspaces
$\cG_1=\mathop{\oplus}\limits_{i\in\Z^N}\cH_1^{(i)}$ and
$\cG_2=\mathop{\oplus}\limits_{i\in\Z^N}\cH_2^{(i)}$, with
$\cG_1\oplus\cG_2=\cG$, represent pure nuclear and pure
molecular channels, respectively.

The Hamiltonian is defined in $\cG$ by the expression
\begin{equation}
\label{MultDimLatHam}
(Hu)^{(i)}=Au^{(i)}+\sum\limits_{j\in\Z^n,\,j\neq i}W(i,j)u^{(j)}
\end{equation}
where the interaction matrices
$$
W(i,j)=\left(\begin{array}{ccc}
       {\sl w}_{11}(i,j) & & {\sl w}_{12}(i,j) \\
       {\sl w}_{21}(i,j) & & {\sl w}_{22}(i,j)
\end{array}\right)\,
$$
consist of the block components ${\sl w}_{mn}(i,j)$ providing
the mappings $\cH_n\rightarrow\cH_m$, $m,n=1,2.$ These components
describe the direct interaction between the $m$-th and $n$-th
channels of the different cells $i$ and $j$, respectively. The
matrices $W(i,j)$ are assumed to be bounded operators in $\cH$
which depend only on the difference
$i-j=(i_1-j_1,i_2-j_2,\ldots,i_N-j_N)$, i.\,e.
\mbox{$W(i,j)=W(i-j)$} and, thus, the same holds for the block
components, \mbox{${\sl w}_{mn}(i,j)={\sl w}_{mn}(i-j)$}.
Moreover, the series of $W(j)$ is assumed to be convergent with
respect to the operator norm topology, i.\,e.
\begin{equation}
\label{WConvergence}
\displaystyle\sum_{j\in\Z_N,\,j\neq 0}\|W(j)\|<+\infty\,
\end{equation}
and the property
\begin{equation}
\label{WSymm}
W(j-i)=[W(i-j)]^*
\end{equation}
is assumed. With such $W(i,j)$ the Hamiltonian
(\ref{MultDimLatHam}) is a selfadjoint operator on the domain
\mbox{$\cD(H)=\mathop{\oplus}\limits_{i\in\Z^N}\cD^{(i)}$} with
$\cD^{(i)}=\cD(h_1)\oplus\C$.  Note that, since $\cH_2=\C$, the
quantities ${\sl w}_{22}(i-j)$ are complex numbers.
The ${\sl w}_{12}(i-j)$ are vectors in
$\cH_1$, the ${\sl w}_{21}(i-j)$ are continuous linear
forms on $\cH_1$, and the ${\sl w}_{11}(i-j)$ are bounded operators
in $\cH_1$.

Let us denote by $\T^N$ the Cartesian product
$\T^N=\underbrace{\T\times\T\times\ldots\times\T}_N$ of the $N$
intervals $\T=[-\pi,\pi]$ and by $p$ the points of $\T^N$,
$p=(p_1,p_2,...,p_N)$, $p_k\in\T$.  The
assumption~(\ref{WConvergence}) implies that the operator-valued
function
\begin{equation}
\label{OmegaP}
\Omega(p)=
\sum_{j\in\Z_n,\,j\neq 0}W(j)\,\exp(\ri\lal p,j\ral),
\qquad \Omega(p):\,\cH\rightarrow\cH, \qquad p\in\T^N,
\end{equation}
where $\lal p,j\ral=\sum_{k=1}^N p_k j_k$
is continuous and bounded on
$\T^N$.  Due to Eq.~(\ref{WSymm}), the values
$$
\Omega(p)=\left(\begin{array}{ccc}
\omega_{11}(p) & & \omega_{12}(p)  \\
\omega_{21}(p) & & \omega_{22}(p)
\end{array}\right)
$$
of this function represent selfadjoint
operators in $\cH$ for any $p\in\T^N$, with
$$
\omega_{mn}(p):\,\cH_n\to\cH_m\,; \quad
[\omega_{11}(p)]^*=\omega_{11}(p), \quad \omega_{22}(p)\in\R
\quad\mbox{and}\quad [\omega_{21}(p)]^*=\omega_{12}(p).
$$
The quantity \mbox{$\widetilde{b}(p)=\omega_{12}(p)$} can be
considered as a vector of $\cH_1$ while
\mbox{$\omega_{21}(p)=\lal\,\cdot,\,\widetilde{b}(p)\ral$}
(cf.~definition~(\ref{H2}) of the Hamiltonian $A$).

The blocks $R(j,k;z)$, $R(j,k;z):\,\cH^{(k)}\to\cH^{(j)}$,
$j,k\in\Z^N$, of the resolvent $R(z)=(H-z)^{-1}$
satisfy the equation
\begin{equation}
\label{MultDimLatRes}
(A-z)R(j,k;z)+\sum\limits_{j'\in\Z_N,\,j'\neq j}
W(j-j')R(j',k;z)=\delta_{jk}I.
\end{equation}
After Fourier transformation in $\cG$,
$$
(Fu)(p)=\frac{1}{(2\pi)^{N/2}}\sum_{j\in\Z^N}
u^{(j)}\exp(\ri\lal p,j\ral),
\qquad p\in\T^N,
$$
the system~(\ref{MultDimLatRes}) takes the form
\begin{equation}
\label{RppMult}
(A-z)R(p,p';z)+\Omega(p)\,R(p,p';z)=\delta(p-p')\,I\,
\end{equation}
where the quasi-momenta $p,p'$ run through the set $\T^N$ and
$R(p,p';z)$ stands for the transformed
resolvent $R(z)$. Thus, the factorization~(\ref{RFactor})
holds with
\begin{equation}
\label{GpzGeneral}
G(p;z)=[A+\Omega(p)-z]^{-1}.
\end{equation}

First, let us consider the specific case
where intercellular interactions
$W(i-j)$ in the Hamiltonian~(\ref{MultDimLatHam})
have the simple form
\begin{equation}
\label{InterCellSimpl}
W(i-j)=\left(\begin{array}{ccc}
       0              &\quad & 0 \\
       0              & & {\sl w}_{22}(i-j)
\end{array}\right)\,,
\end{equation}
i.\,e., where the
cells interact with each other only via the molecular channels.
Obviously, in this case the factor $G(p;z)$ is still given by
(\ref{Gpz}). As compared to this expression the
only difference is that now $p\in\T^N$ and
\begin{equation}
\label{M2tildMult}
\tM_2(p;z)=\lambda_2-z+\omega_{22}(p)-\beta(z)\,.
\end{equation}

Let $\omega_{22}^{\rm min}=\mathop{\rm min}\limits_{p\in\T^N}
\omega_{22}(p)$ and $\omega_{22}^{\rm max}=\mathop{\rm
max}\limits_{p\in\T^N} \omega_{22}(p)$.  Similarly to the
analogous assumption in Sec.~\ref{Lattice1Dim} we assume that
the closed interval \mbox{$[\lambda_2+\omega_{22}^{\rm
min},\lambda_2+\omega_{22}^{\rm max}]$} is totally embedded in
the absolutely continuous spectrum $\sigma_c(h_1)$ of $h_1$, and
no thresholds of this spectrum belong to
\mbox{$[\lambda_2+\omega_{22}^{\rm
min},\lambda_2+\omega_{22}^{\rm max}]$}.  We also assume that
this interval belongs to the holomorphy domain $\cD$ of the
function $\beta(z)$, and that for $\mu\in[\lambda_2+\omega_{22}^{\rm
min},\lambda_2+\omega_{22}^{\rm max}]$ the
inequality~(\ref{NeverZero}) holds.

Let us consider
the time evolution of the system
described by the Hamiltonian~(\ref{MultDimLatHam}) with the
simple intercellular interactions~(\ref{InterCellSimpl})
subject to the above conditions. We
start again from a pure molecular state
$\varphi=\varphi_1\oplus\varphi_2$, $\|\varphi_n\|\in\cG_n$,
$n=1,2$, with $\varphi_1=0$ and $\|\varphi\|=\|\varphi_2\|=1$.
The probability $P_{\rm mol}(\varphi,t)$ to find the system at a
time  $t\geq0$  in the molecular channel is given by
the analogue of (\ref{Pphi}).  As in
Sec.~\ref{Lattice1Dim} one finds the quasi-momentum
representation
$$
   \biggl(\sP_2 R(z)\varphi\biggr)(p)=G_{22}(p;z)\varphi_2(p)=
    \frac{1}{\tM_2(p;z)}\,\varphi_2(p)\,,\qquad p\in\T^N\,,
$$
and relations (\ref{P2phiFactor}),
(\ref{Jpt}) and~(\ref{JptBehavior}) are still valid with the
only difference that instead of~(\ref{lambdatilde})
$\widetilde{\lambda}_2(p)$ is now of the form
$$
   \widetilde{\lambda}_2(p)=\lambda_2+\omega_{22}(p)\,,\quad p\in\T^N\,.
$$

According to Eqs.~(\ref{zNucl}) and~(\ref{zMol})
the main terms of the roots
$z_{\rm nucl}(p)$ and $z_{\rm mol}(p)$ of
the function~(\ref{M2tildMult}) in the unphysical sheet read
\begin{eqnarray}
\label{zNuclpMult}
 z_{\rm nucl}(p)  & \cong &
z_1-\D\frac{a}{ \lambda_2+\omega_{22}(p)-z_1 },\\
\label{zMolpMult}
z_{\rm mol}(p) & \cong &
\lambda_2+\omega_{22}(p)+\D\frac{a}{\lambda_2+\omega_{22}(p)-z_1}.
\end{eqnarray}

The asymptotic equation (\ref{JptBehavior}) now implies
\begin{equation}
\label{IntExpGammaMult}
P_{\rm mol}(\varphi,t)=\int_{\T^N}
dp\,|J(p,t)|^2\,|\varphi_2(p)|^2=
\int_{\T^N} dp\,
\exp\{-{\Gamma_R^{(m)}(p)}\,t\}\,|\varphi_2(p)|^2
+\widetilde{\varepsilon}(t)\,
\end{equation}
with
$$
\Gamma_R^{(m)}(p)=
-2\Img z_{\rm mol}(p)\cong
-2\Img\D\frac{a}{\lambda_2+\omega_{22}(p)-z_1}.
$$
As in Sec.~\ref{Lattice1Dim} the background term
$\widetilde{\varepsilon}(t)$ is small for any $t\geq 0$,
$\widetilde{\varepsilon}(t)=O(\|b\|^2)$ and
$|\widetilde{\varepsilon}(t)|\ll1$.

Thus, if the real part $E_R^{(1)}$ of the ``nuclear'' resonance
$z_1$ belongs to the interval \mbox{$[\lambda_2+\omega_{22}^{\rm
min}, \lambda_2+\omega_{22}^{\rm max}]$} then there are
``molecular'' states $\varphi$ which decay via the ``nuclear''
channel with a rate as close as possible to the maximal
value~(\ref{GmFin}).  In this case the components $\varphi_2(p)$ are
localized in a close neighborhood of the manifold
\begin{equation}
\label{Manifold}
   \lambda_2+\omega_{22}(p)=E_R^{(1)}\,
\end{equation}
in the quasi-momentum space $\T^N$.
In particular, if the initial state $\varphi$ is prepared such
that the component $\varphi_2(p)$ is nonzero only for the
quasi-momenta $p$ lying in the domain
$|\lambda_2+\omega_{22}(p)-E_R^{(1)}|\leq\delta\,\Gamma_R^{(1)}/2$
with some small $\delta>0$,  then one should only integrate over
this domain in the integral~(\ref{IntExpGammaMult}). In such a
case, under the condition~(\ref{ReaIma}),
a lower estimate for the decay rate $\Gamma_R^{(m)}$
is given, as in Sec.~\ref{Lattice1Dim}, by
$\D\frac{1}{1+\delta^2}\,\D\frac{4\Real a}{\Gamma_R^{(1)}}$.
Thus, by varying $\delta$ one can
get a rate as close as possible to the
maximum~(\ref{GmFin}).

Now, let us consider briefly the case of more general intercellular
interactions $W(i-j)$ where all the components $\omega_{mn}(p)$,
$m,n=1,2,$ of the matrix $\Omega(p)$, $p\in\T^N$, can be
nontrivial.  In this case the component $G_{22}(p;z)$ of the
factor~(\ref{GpzGeneral}) is given again by Eqs.~(\ref{G22})
and~(\ref{M2tildMult}). However, the function $\beta(z)$
in~(\ref{M2tildMult}) is to be replaced by the modified
function
$$
   \widetilde{\beta}(p;z)=\lal[h_1+\omega_{11}(p)-z]^{-1}
   [b+\widetilde{b}(p)],[b+\widetilde{b}(p)]\ral\,
$$
where \mbox{$\widetilde{b}(p)=\omega_{12}(p)$}.
We make the natural assumption that
the direct intercellular interactions in nuclear channels
${\sl w}_{11}(i-j)$ are so weak that the term $\omega_{11}(p)$
produces only a very small perturbation of the initial
``nuclear'' resonance $z_1$ generated by the Hamiltonian
$h_1$. More precisely, we assume that the resonance
$\widetilde{z}_1(p)$ generated by the perturbed Hamiltonian
$h_1+\omega_{11}(p)$ has the property
\begin{equation}
\label{z1tilde}
  |\widetilde{z}_1(p)-z_1|\ll\Gamma_R^{(1)}
  \quad\mbox{\rm for any}\quad p\in\T^N\,
\end{equation}
and that no other resonances arise in the domain $\cD$.  Another
natural assumption is that the strength
of the interactions $\sw_{12}(i-j)$ and $\sw_{21}(i-j)$, $i\neq
j$, between ``nuclear'' and ``molecular'' channels of
different cells is much weaker than the one of a
single cell. This is why we can assume
\mbox{${\|\widetilde{b}(p)\|}/{\|b\|} \lesssim 1$}
and the following features.
The elements $\widetilde{b}(p)\in\cH_1$ are such that for any
$p\in\T^n$ the function $\widetilde{\beta}(p;z)$ admits an analytic
continuation into the domain $\cD$ of the unphysical sheet
and in $\cD$ a representation of the type~(\ref{BetaRepres})
holds for $\widetilde{\beta}(p;z)$,
\begin{equation}
\label{BetaRepresP}
\widetilde{\beta}(p;z)=
\D\frac{\widetilde{a}(p)}{\widetilde{z}_1(p)-z}+
\widetilde{\beta}^{\rm reg}(p;z),
\end{equation}
with an explicitly separated pole term
$\widetilde{a}(p)/(\widetilde{z}_1(p)-z)$ and a holomorphic
remainder $\widetilde{\beta}^{\rm reg}(p;z)$.  As in
Sec.~\ref{Perturbation} we assume that
$|\widetilde{a}(p)|=C_a(p)\|b\|^2$, and that for any $p\in\T^N$ the
limiting procedure (\ref{alimit}) is possible for $C_a(p)$
while $\Img\widetilde{a}(p)\ll\Real\widetilde{a}(p)$. For the
remainder $\widetilde{\beta}^{\rm reg}(p;z)$, \mbox{$p\in\T^N$,}
\mbox{$z\in\cD$,} the same statements are assumed to be valid as for the
function $\beta^{\rm reg}(z)$.  Under these assumptions one can
repeat almost literally the study of the probability $P_{\rm
mol}(\varphi,t)$ as performed above in
case of the interactions~(\ref{InterCellSimpl}). We
find again that the asymptotics of $P_{\rm mol}(\varphi,t)$ is
given by (\ref{IntExpGammaMult}) with
$$
\Gamma_R^{(m)}(p)\cong
-2\Img\D\frac{\widetilde{a}(p)}{\lambda_2+\omega_{22}(p)
-\widetilde{z}_1(p)}\,.
$$
Let us denote by $(\Real \widetilde{a})_{\rm max}$ the maximal
value of the function $\Real \widetilde{a}(p)$ on the manifold
(\ref{Manifold}). It is obvious that if one prepares the initial
pure``molecular'' state $\varphi$ such that its
component $\varphi_2(p)$ is localized in a close neighborhood of the
subset of the manifold (\ref{Manifold}), where $\Real
\widetilde{a}(p)=(\Real \widetilde{a})_{\rm max}$, then for the
probability $P_{\rm mol}(\varphi,t)$ the main qualitative result
remains practically the same as in case of the
interactions~(\ref{InterCellSimpl}). Namely, varying the support
of the component $\varphi_2(p)$ in $\T^N$ one can achieve a decay
rate of the state $\varphi$ as close as possible to the maximal
value of the width $\Gamma_R^{(m)}(p)$ in (\ref{Manifold}). The
main term $4(\Real \widetilde{a})_{\rm max}/\Gamma_R^{(1)}$ of
this value is again inversely proportional to the ``nuclear''
width $\Gamma_R^{(1)}$.


\bigskip
\acknowledgements
One of the authors (A.\,K.\,M.) is much indebted to
Prof.~W.\,Sandhas for his hospitality at the Universitaet Bonn.
Support of this work by the Deutsche Forschungsgemeinschaft
is gratefully acknowledged. This work was partially supported
also by the Russian Foundation for Basic Research.


\end{document}